\documentclass{article}
\usepackage{PRIMEarxiv}

\usepackage[utf8]{inputenc} 
\usepackage[OT2, T1]{fontenc}    
\usepackage{hyperref}       
\usepackage{url}            
\usepackage{booktabs}       
\usepackage{amsfonts}       
\usepackage{nicefrac}       
\usepackage{microtype}      
\usepackage{lipsum}
\usepackage{fancyhdr}       
\usepackage{graphicx}       
\graphicspath{{media/}}     
\usepackage{textgreek}
\usepackage{cite}

\title{In Silico Tools in PROTACs Design
}

\author{
  Mengman Wei \\
  \texttt{{weimengman}77@gmail.com}\\
}

\begin{document}
\maketitle

\begin{abstract}
{PROTACs (PROteolysis TArgeting Chimeras), as a highly promising new therapeutic paradigm, have attracted widespread attention from the academic and pharmaceutical communities in recent years. To date, the design and validation of PROTACs molecule’s druggability primarily rely on experimental approaches, making the development process of this kind of drug molecule time-consuming. Computer-aided tools for PROTACs design may offer a potential solution to expedite the design process and enhance its efficiency. This mini review briefly summarizes the in silico tools for PROTACs drug molecule design reported recently.}
\end{abstract}


\section{Introduction}
\begin{figure}
    \centering
    \includegraphics{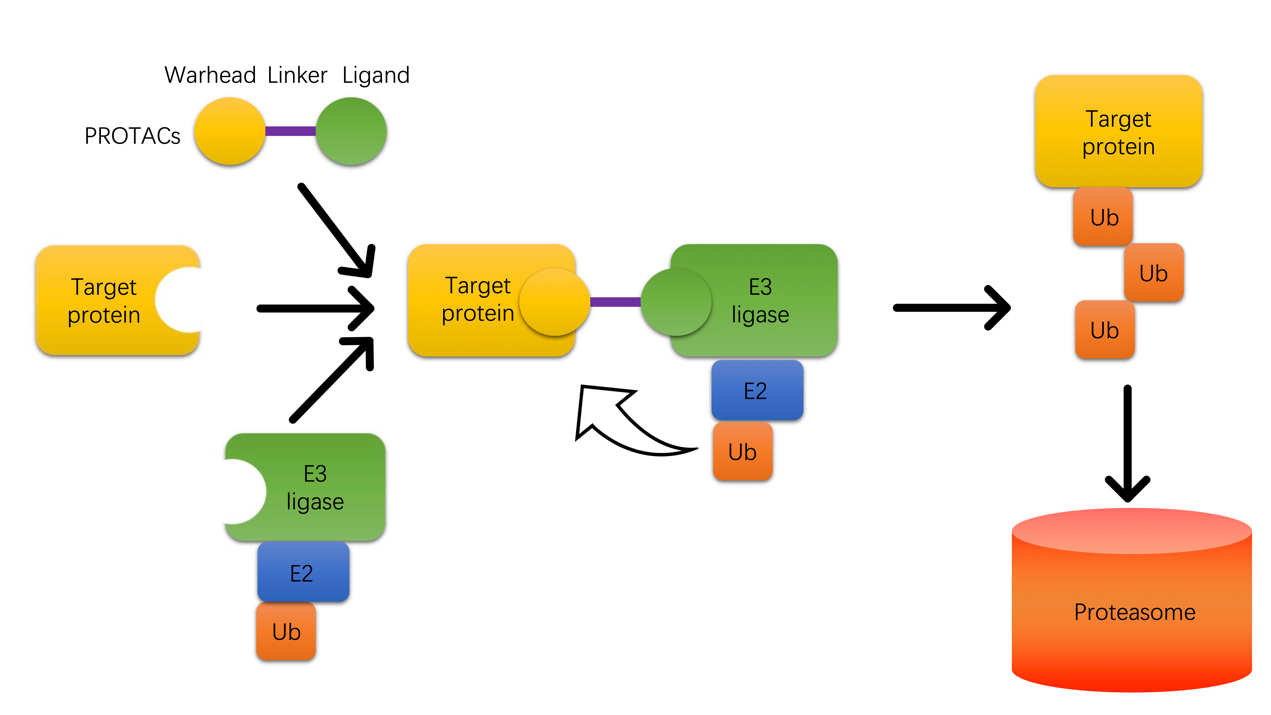}
    \caption{A simple diagram shows how a PROTACs molecule utilizes the ubiquitin-proteasome system(UPS) to degrade a target protein}
    \label{f1}
\end{figure}

{PROTACs (PROteolysis TArgeting Chimeras) are a class of molecules that harness the body's natural protein degradation machinery to selectively remove disease-causing proteins. PROTACs consist of three components: a warhead that binds to the target protein, a ligand that binds to an E3 ubiquitin ligase, and a linker connecting these two components. This molecule makes the target protein and the E3 ligase proximity, leading to the transfer of ubiquitin molecules from the E3 ligase to the target protein. These ubiquitination tags draw proteasome to degrade the target protein (Figure ~\ref{f1}).}

{Compared to traditional small-molecule therapeutics, PROTACs exhibit two major advantages. First, traditional small-molecule therapeutics primarily function through the identification and screening of molecules capable of inhibiting proteins of interest (POIs), employing an occupancy-based inhibitory mechanism that necessitates the identification of a suitable binding pocket on the POIs. However, it is estimated that approximately 80\% of POIs present challenges in identifying amenable binding pockets, rendering them categorized as “undruggable” targets \cite{1, 2, 3}. PROTACs, on the other hand, operate through a different paradigm by promoting the degradation of pathogenic proteins, thus offering significant potential to surmount the challenges associated with target undruggability. Second, the occupancy-based inhibitory mechanism of traditional small-molecule therapeutics necessitates sustained high concentrations of drug exposure, which may give rise to adverse effects \cite{2, 4}. Additionally, binding pockets are susceptible to point mutations which can culminate in the emergence of drug resistance upon prolonged treatment regimens \cite{5, 6, 7, 8}. PROTACs, which are not contingent upon binding pocket interactions, represent a more versatile approach in this regard. Furthermore, for other emerging promising therapies, such as RNA interference and antibody-based therapies, PROTACs also demonstrate some advantages. Compared to the off-target effects on gene level that RNA interfering might cause, which leads to serious undesired effects \cite{4, 9, 10, 11}, PROTACs, functioning at the post-translational level, inherently circumvent these risks. And antibody therapy still has limitations as it cannot cross the cell membrane and requires extravascular administration\cite{8}.}

{Following the elucidation of the functional mechanisms underpinning the ubiquitin-proteasome system (UPS) \cite{12, 13, 14}, scientific endeavors were initiated to harness the UPS for targeted protein degradation \cite{15, 16}. The PROTACs technology emerged as an innovative approach that simulates natural processes \cite{17}. Initially coined as a “chemical curiosity” by the Crews and Deshaies teams \cite{17}, PROTACs underwent a series of rigorous investigations, starting with in vitro assays and cellular environmental assays \cite{17, 18, 19, 20, 21, 22, 23, 24, 25, 26, 27, 28, 29, 30, 31, 32, 33, 34, 35, 36, 37, 38, 39, 40, 41, 42, 43, 44, 45, 46, 47, 48, 49, 50, 51}, advancing to in vivo animal model studies \cite{52, 53, 54, 55, 56, 57, 58, 59}, and eventually making the leap from the bench to clinical evaluations \cite{60, 61, 62, 63, 64, 65}. This trajectory has encompassed a systematic progression, fulfilling the critical milestones of concept validation across cellular, animal, and human models. A landmark achievement was attained in 2019 when Arvinas’ PROTAC candidate, ARV-110, successfully completed Phase I clinical trials, marking the first clinical evidence of the feasibility of PROTACs in intervening in human diseases \cite{60, 61, 62}. PROTACs therapeutic technology continues to advance and mature through the solid research of various scientific teams in this field. Both laboratory and clinical data have demonstrated the great possibility of PROTACs as a novel therapeutic approach. After 20 years of validation and development, this field has gradually emerged as a therapeutic paradigm with immense potential.}

{Current PROTACs related research and development are mainly based on trials, which could be time-consuming. In silico tools may assist researchers in delving deeper into the details of how PROTACs function on the atom level, thereby helping them to outline strategies for designing such molecules. Here this paper tries to summarize computer-aided PROTAC design tools that have been developed in recent years, to see if these methods can offer some insights into the design process.}

\begin{figure}
    \centering
    \includegraphics{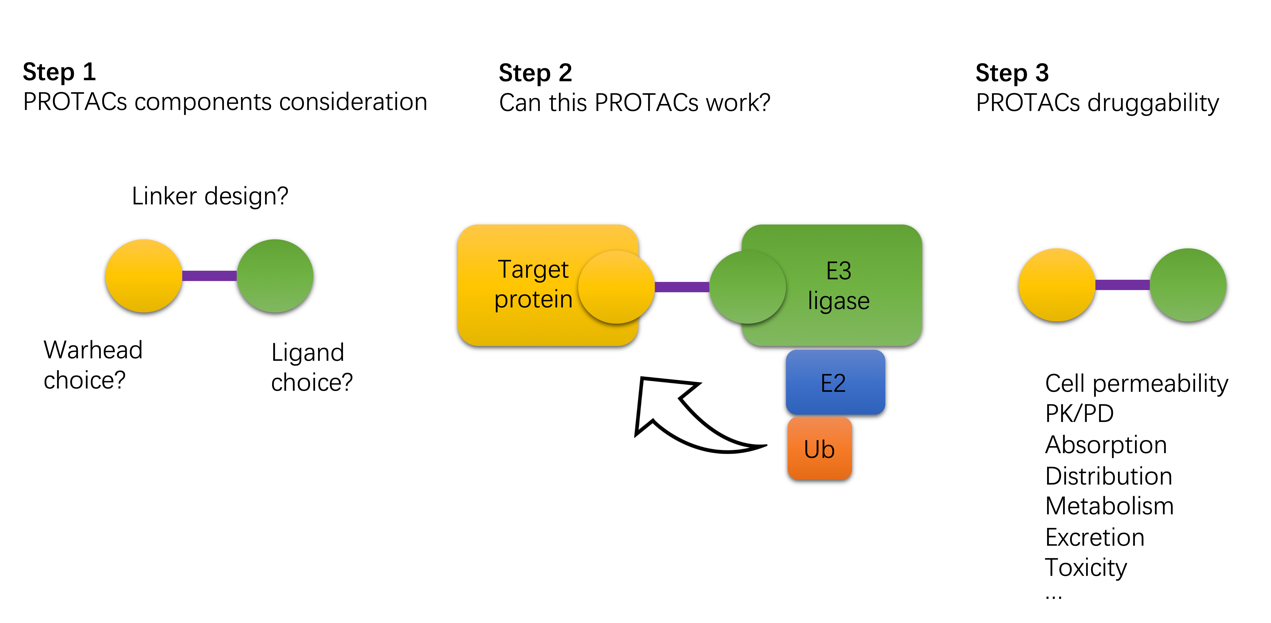}
    \caption{Three main consideration steps in computer-aided PROTACs design}
    \label{f2}
\end{figure}

{The general design of PROTACs strategy includes the selection of molecular components, the prediction and evaluation of the molecule's efficacy, as well as an evaluation of its pharmacological attributes in the context of druggability (Figure ~\ref{f2}). Following this order, subsequent sections will delineate the computational methodologies reported in the literature.}

\section{Design and selection of each component of PROTACs}
{Currently, the selection of the warhead targeting the POIs does not present a challenging task. The principal criterion for a warhead is its binding affinity toward the target protein. This contrasts with the traditional approach where, beyond affinity, molecules are subjected to extensive evaluation for aspects such as selectivity, to mitigate the risk of off-target effects. In the context of PROTACs, the decision-making for warhead selection is streamlined, with affinity being the solitary consideration. Moreover, there exists plenty of tools that have been developed within the ambit of traditional small-molecule drug discovery paradigms, aimed at isolating molecules with affinity to target proteins \cite{66, 67, 68, 69, 70, 71, 72, 73, 74, 75, 76, 77, 78, 79, 80, 81, 82, 83, 84, 85, 86, 87, 88, 89, 90, 91, 92}. These tools can be leveraged for the efficacious screening of warheads for PROTACs. Interestingly, most of the warheads employed in clinical PROTACs exploit existing inhibitors of target proteins \cite{60, 64}. A noteworthy merit of this approach is the potential therapeutic effect through the inhibition of disease-causing proteins, in instances where PROTACs may not achieve the intended protein degradation.}

{The rationale employed in the design of ligands for E3 ligase mirrors that utilized for the selection of warheads targeting POIs, with binding affinity being a central consideration in both instances. However, it is imperative to highlight a distinguishing feature; whereas the objective with POIs is to achieve complete degradation, E3 ligases must retain their biological functionality, including the transfer of ubiquitin and re-entry into the UPS degradation cycle. Consequently, after the identification of ligands exhibiting high affinity, it is essential to ascertain that these ligands do not compromise the intrinsic functions of the E3 ligase.}

\begin{figure}
    \centering
    \includegraphics{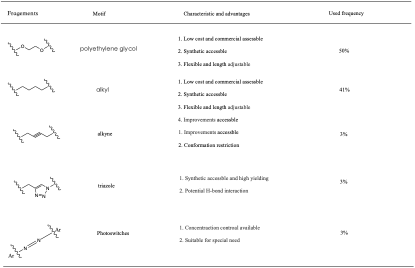}
    \caption{Selected linker motifs reported so far \cite{94}}
    \label{f3}
\end{figure}

{A critical obstacle in the design of PROTACs pertains to the selection of an appropriate linker, and studies have suggested that the constitution and length of the linker exert a significant impact on the efficacy of the targeted degradation process \cite{51, 93}. A review of linkers employed in PROTACs development reveals a predominance of simple PEG or alkyl chains \cite{94} (Figure ~\ref{f3}). Notably, the design of linkers has been predominantly empirical, guided by experiential knowledge rather than established principles. To date, there is no definitive understanding regarding the attributes of linker substrates or lengths that might enhance the degradation process. For example, Zorba et al. once designed a series of PROTACs with linkers of varying lengths to degrade the BTK. Under the PROTAC structures they designed, PROTACs with linker lengths of around 20 atoms demonstrated better efficacy \cite{95}. However, in other PROTAC design studies, it has been found that certain PROTAC molecules exhibit optimal efficacy when the linker length is only 3 atoms \cite{96}. Given the lack of universal guidelines, the selection of linkers in PROTACs design necessitates a tailored approach, accounting for the particularities of each case.}

{For the generation algorithm of the linker, novelty and effectiveness are usually considered. Most molecular generation algorithms will produce datasets with property distributions similar to the training set \cite{97, 98, 99, 100, 101, 102, 103, 104, 105, 106, 107}, and they are good for general molecular generation tasks, but this approach is not applicable to the generation of PROTAC linkers, because the current PROTAC linkers dataset is very small and lacks diversity. The way to generate PROTAC linkers should be able to help us break through the boundaries of a chemist’s knowledge and generate linkers that are innovative and within the drug-likeness space. There was a report of such work in 2020, named DeLinker \cite{108}.}

{DeLinker is a graph-based deep generative model that combines with 3D structural knowledge\cite{108}. They used ZINC \cite{109} and CASF \cite{110} as training datasets. The model tried to get 3D structures first, the 3D structures are gotten by the lowest energy 3D conformers for the ZINC dataset using the RDKit \cite{111} tool and experimentally verified active ones from CASF dataset. And then they cut all the molecules in the dataset into two components and a linker. The molecular structure information is represented in a graph format where atoms can be represented as nodes and bonds as edges. The 3D information includes the relative distance and orientation between the fragments or substructures. Then the model tried to learn how each pair in the training set are connected by a linker. It would be trained to generate linkers that adhere to these constraints and are chemically feasible capturing complex patterns in the data. Once it is trained, it could be used to generate a linker for any given two parts that needed to link together. The model employs an iterative approach to construct the linker by progressively adding bonds from a repository of atoms, which can be initialized with partial structures. The user has the discretion to determine the maximum length of the linker. The selection of nodes is systematic and adheres to a deterministic first-in-first-out queue, initialized with the exit vectors of each fragment. When a node becomes part of the graph for the first time, it is queued for consideration. The model continues to append edges to the node in question until a termination criterion is satisfied, indicating the completion of that particular segment. Once this state is reached, the node is designated as ‘closed,’ precluding any further addition of edges. In the decision-making process for appending an edge, the model exhibits remarkable discernment by integrating a wealth of information. It considers the immediate attributes of the nodes (local information), takes into account the overall structural framework including the unlinked fragments (global information), and crucially incorporates 3D spatial data. The incorporation of this 3D structural information is indispensable and the researchers have proved this in their paper. This model is able to generate novel linkers for PROTACs that bear resemblance to the molecules found in ZINC or CASF databases in terms of properties. Notably, the molecules in these databases have already been proven to be biocompatible with the human body. Furthermore, this tool offers researchers the capability to design linkers, contingent upon the acquired interaction structures between target proteins and E3 ligases. These 3D protein-protein interaction structures can be obtained via molecular dynamics simulations or through actual crystallographic data. Utilizing this approach enhances the success rate of creating linkers that are tailored to these protein-protein interactions (PPIs), which are the dominance of the interaction network within the target protein-PROTACs-E3 ligase ternary complex.}

{Apart from DeLinker, there are several machine learning-based linker generation algorithms that should help with PORTACs linker design reported. DiffLinker \cite{112}, a recently introduced E(3)-equivariant 3D-conditional diffusion model, has been reported to be adaptable based on protein pockets and can generate linkers with minimal clashes. Notably, DiffLinker transcends prior fragment-linking methods by being able to connect more than just pairs of fragments, scaling naturally to an arbitrary number of them. While the study did not specifically explore its utility in PROTACs design, it appears to hold promise for such an application. Furthermore, two reinforcement learning-based methods for PROTACs linker generation, namely Link-INVENT \cite{113} and DRLinker \cite{114}, were published. These models are capable of generating linkers with desired properties by employing reinforcement learning in conjunction with a specified scoring function. They have been validated to be effective for various fragment-linking design objectives if certain properties are sought after. It is pertinent to mention that the selection criteria for PROTACs linkers are not well-defined yet and it might require a scientist with a lot of experience in PROTACs development to guide the linker generation process when using these tools.}

\section{Prediction and evaluation of the degradation efficacy of PROTACs }

{After the selection of the three constituents of PROTACs, the subsequent step involves ascertaining the rationality of the molecule's design and evaluating its accessibility for chemical synthesis. There exists a repertoire of well-established tools that are user-friendly and aid researchers in analyzing the complexity involved in synthesizing a molecule \cite{115, 116, 117, 118}.}

{Another critical aspect to evaluate at this stage is the functionality of the PROTACs - whether they are effective and capable of degrading the target protein. The majority of efficacy assessment methods established to date focus on ascertaining the stability of the target protein-PROTACs-E3 ligase complex that is formed. Ciulli et al. \cite{51} found that PROTAC molecules achieve their function by properly folding and bending through intermediate linker chains, allowing both ends to penetrate into the hydrophobic cavities of the two receptors, bringing E3 ubiquitin ligase and the target protein closer to each other. This forms specific target protein-PROTACs-E3 ligase PPIs, promoting the ubiquitination of the target protein. It is then believed that stable PPIs can facilitate the degradation of the target protein.}

{In earlier years, there were some computational studies involving docking/dynamics simulations of the ternary complex consisting of target protein-PROTACs-E3 ligase \cite{51, 54, 95, 119, 120, 121, 122, 123, 124, 125}. These SAR analyses were combined with actual experimental data to investigate the relationship between the structure of these complexes and properties such as the selectivity/degradation efficiency of designed PROTAC molecules. }

{In 2019, Drummond et al. reported a pioneering work on establishing a universal modeling PROTACs mediated ternary complexes strategies based on the MOE (Molecular Operating Environment) \cite{126}. This model is adept at predicting the potential structure of ternary complexes for a given combination of target protein, PROTACs, and E3 ligase, which is instrumental in further assessing the predicted structures in terms of their likely degradation effects. They investigated four distinct modeling strategies for the complexes: Method 1 entails simultaneous sampling of the entire ternary complex. Method 2 involves the independent sampling of PROTACs conformations, followed by the subsequent integration of rigid-body proteins. Method 3 comprises sampling the PROTACs in relation to one of the proteins and then incorporating the second protein. Method 4 involves the independent sampling of PROTACs conformations without the proteins (Method 2 can also contribute to the conformational database or the other way around), while protein-protein docking guides possible ligase-target arrangements. Among these methods, Method 4 stood out in performance and was successful in closely mirroring the crystal structures. The efficacy of Method 4 was enhanced by integrating a clustering process that streamlined the final output ensemble of PROTAC-mediated ternary complexes \cite{127}.}

{Several additional modeling methods have been reported. Primarily, these methods extend on strategy 4 outlined by Drummond et al. \cite{126}, which entails initial protein-protein docking succeeded by the incorporation of the PROTAC molecule. These methods enhance and fine-tune the scoring and clustering processes at each stage of strategy 4, albeit employing diverse programming languages or computational packages. For instance, PRosettaC \cite{128}, which is built upon the Rosetta suite, executes Local Docking Refinement post-protein docking, generating 50 high-resolution models for every protein-protein global docking configuration. This approach enables the protocol to identify native solutions. Furthermore, they offer an accessible web server that is tailored to facilitate usage by those who are novices in computer-aided drug design tools. Another technique, also based on Rosetta \cite{129}, integrates the linker into the protein-protein complex and carries out energy minimization filtering. Given that the protein-attracting segments of the PROTAC remain static during the PROTAC assembly, the relative position of the PROTAC with respect to the protein components is known. This facilitates the straightforward construction of a comprehensive ternary complex model, which is then subject to energy minimization within Rosetta. Post filtering, the Fraction of Fully Compatible Complexes (FFC) is computed, evaluating the geometric and energetic compatibility of a given PROTAC linker with the protein pair’s favored interaction modes. Ultimately, FFC is formulated to assess the alignment between the PROTAC-imposed constraints and the intrinsic preferences apparent in the initial ensemble of docked models. Their results demonstrated that it is possible to retrospectively explain relationships between linker length/composition and cellular activity. Most recently, PROTAC-Model \cite{130}, an all-encompassing computational protocol, has emerged. They used FRODOCK \cite{131} for confined local docking, RosettaDock \cite{132} for structural refinement, RDKit \cite{111} for modeling PROTAC conformations, Open Babel (Obenergy) \cite{133} for PROTAC conformation evaluation, AutoDock Vina \cite{134} for PROTAC binding mode assessment, and VoroMQA \cite{135} for reranking protein-protein complexes. The protocol has demonstrated superior efficacy in modeling near-native ternary complex structures starting from unbound structures.}

{The aforementioned modeling methods have shown competence in exploring the conformational space as the native pose is usually present among the set of poses generated. However, there is inconsistency in the ability of the pose filtering or scoring approaches to reliably identified the native pose in different systems. Observing that the native crystal ternary pose remains in a stable state during molecular dynamics (MD) simulations of up to around 100 ns, while non-native poses may not necessarily maintain stability, Tang et al. \cite{136} devised a method that employs HAPOD trials of candidate poses coupled with explicit solvent MD to assess the average pose occupancy fraction under gradual heating. Initial candidate poses were generated via MD sampling and preliminarily ranked using classic MM/GBSA. A specialized heating protocol was then implemented to hasten the departure of the ternary pose while tracking the pose occupancy duration and fraction. The HAPOD scoring method succeeded in identifying the native pose in all tested systems. This research illustrates that native and non-native ternary complex poses can be differentiated based on pose occupancy duration in MD, with native poses exhibiting extended occupancy times at both room and higher temperatures. The success of this method is in part due to the dynamic assessment of pose departure, which considers entropic effects often overlooked by static scoring techniques. As entropy has a more pronounced role in protein-protein interactions compared to protein-ligand interactions. This study offers a solid and reliable way to evaluate the stability of the PROTAC-induced ternary complex.}

{In recent years, several data-driven methods based on machine learning model for evaluating the effectiveness of PROTACs have been reported.}

{Yang and his team developed a PROTAC design platform based on machine learning methods \cite{137}. The model takes an E3 ligand and a warhead as inputs and creates customized linkers, leading to the creation of PROTACs that are chemically sound and have good qualities. They started by training a model using a neural network and a large dataset of molecules similar to PROTACs because there wasn't enough PROTAC data. They improved this model using real PROTACs and random SMILES representations of fragments. They combined this improved model, called the Performer, with a memory-boosted Reinforcement Learning system to help create PROTACs with better drug-like properties. As a proof of concept, they made 5,000 PROTACs targeting a protein called BRD4 to test the platform. After sorting and testing them with machine learning and simulations, they made six of these PROTACs and tested them in the lab. Three of them were effective against BRD4. One of them was particularly good at stopping the growth of Molt4 cells and worked well in mice.}

{A deep neural network model to predict the degradation efficacy of a given PROTACs based on the structures of POI and E3 ligase, named DeepPROTACs, was introduced \cite{138}. Their method employs distinct neural network modules to encode various parts of a given POI-PROTAC-E3 ligase complex. The embeddings of these components are concatenated and input into a Multilayer Perceptron (MLP) comprising two fully connected layers to generate the ultimate output. The model registered an average accuracy rate of 77.95\% and an AUROC (Area Under the Receiver Operating Characteristic) value of 0.8470 in the test set. Moreover, the model underwent validation with a collection of PROTACs that harnessed VHL for targeting the estrogen receptor (ER). Among 16 PROTACs, the model proficiently anticipated the degradation capacities of 11, reaching a prediction accuracy of 68.75\%. }

{As real experimental data on PROTACs burgeons, machine learning methods rooted in mathematical statistics will expedite and enhance the prediction of PROTAC degradation capabilities, showcasing their considerable potential.}

\section{Prediction and evaluation of the druggability of PROTACs}
{Once confirming a PROTAC molecule is functional, it is then worthwhile to consider developing it into a drug. Typically, it will be evaluated based on its physicochemical properties, ADMET (Absorption, Distribution, Metabolism, Excretion, and Toxicity) properties, and so on. There are already many well-established tools to help predict a molecule's physicochemical properties \cite{139, 140, 141, 142, 143, 144, 145}. However, since the size of PROTAC molecules is larger than traditional small drug molecules, some machine learning (ML)-based prediction results may not be as accurate as those for traditional small drug molecules, which are used in the training set. The methods used to calculate metabolism and toxicity for small molecule drugs cannot be directly applied to PROTACs, because they interact with target proteins in different ways. Traditional small molecule drugs work by occupying a site on the target, while PROTACs function by degrading the target protein and then returning to circulate in the bodily fluids. Therefore, there is a need to develop new methods for analyzing the membrane permeability, toxicity, and kinetic properties of PROTACs.}

{The ability of PROTACs to permeate cell membranes seems to be a critical characteristic, as it determines the feasibility of oral administration. Kihlberg et al. \cite{146} employed NMR spectroscopy and molecular dynamics (MD) simulations separately to delve into the factors behind the varying cell permeabilities shown by three structurally similar, flexible cereblon PROTACs. Both analytical techniques revealed a correlation between high cell permeability and the propensity of the PROTACs to assume folded structures with a low exposure of their 3D polar surface area to solvent in a non-polar setting. The chemical composition and adaptability of the linker were found to be vital in allowing the PROTACs to maintain folded structures, which were stabilized by intramolecular hydrogen bonds, Pi-Pi interaction, and van der Waals forces. This study offers a valuable methodology for examining the permeability of PROTACs across cell membranes.}

{Mathematical modeling methods have been found to be used to quantitative analysis of ternary complexes \cite{147, 148}. In 2021, Gilbert and colleagues \cite{149}developed a mechanistic model for Bispecific Protein Degraders (BPDs) that takes into account the reaction pathways involved in the formation of ternary complexes and their subsequent degradation by the ubiquitin-proteasome system (UPS). An essential feature of this model is the inclusion of multiple steps which create a time lag between the formation of the ternary complex and the degradation of the protein. This time lag is crucial as it establishes a balance between the stability of the ternary complex and the rate of degradation by the UPS. This is similar to the concept of kinetic proofreading, which has been suggested to account for the precision and specificity of various biological processes, including protein synthesis and T-cell receptor signaling. Kinetic proofreading is thought to be central to how cells control the recognition and degradation of substrates by the UPS. The model put forth by Gilbert and his team employs this kinetic proofreading concept within a quantitative framework that integrates pharmacokinetics (PK) and pharmacodynamics (PD), providing valuable insights for the development of effective and selective BPDs. These recent advancements, coupled with biophysical investigations, offer a more comprehensive understanding of the relationship between the kinetics of ternary complexes and the efficiency of PROTAC-induced degradation.}

\section{Conclusion}
{Even with the limitation of experimental data, computational methodologies encompassing component selection, efficacy evaluation, membrane permeability analysis, and kinetic analysis for PROTACs have been cultivated to offer insights into PROTACs design. As PROTACs represent a potentially revolutionary approach to therapy, there is a pressing need to forge more substantial in-silico tools for dissecting the structure-activity relationship between PROTACs and target proteins. In doing so, this empowers researchers with an enriched comprehension of PROTACs' biological machinations within the human body, facilitating the distillation of design principles for PROTACs drug molecules, and accelerating research advancements.}

\bibliographystyle{unsrt}  
\bibliography{references}

\end{document}